# Distribution networks and the optimal form of the kidney and lung.

Walton R. Gutierrez*
*Touro College, 27 West 23rd Street, New York, NY 10010*
*Electronic address: waltong@touro.edu

**A model is proposed to minimize the total volume of the main distribution networks of fluids in organs such as the kidney and the lung. A consequence of the minimization analysis is that the optimal overall form of the organs is a modified ellipsoid. The variational procedure implementing this minimization is similar to the traditional isoperimetric theorems of geometry.**     **PREPRINT, May 2005**
   Note: *An expanded version of this article will be available sometime in Dec. of 2005, taking a closer look at the fractal nature of the network.*

At the allometric level there is similarity between the kidney and lung of mammals, when these organs are anatomically described in terms of the dimensions of their organ sites, such as the nephrons and alveoli [1-3].
The kidney and lung are notable for their capacity to process large amount of fluids in a small organ volume, thereby indicating that the distribution networks of fluids of these organs have a geometric design of great efficiency. A basic requirement to achieve this efficiency is the minimization of the total volume of the network, so that a larger part of the volume of the organ is available to the nephrons and alveoli where the actual processing of fluids is performed. Computer simulations of distribution networks of simplified box-like organs already show the large effect of the organ form in the total volume of an efficient distribution network [4].
The model developed here links geometric characteristics of distribution networks to main features of organ anatomy: from the microscopic sites to the macroscopic overall form of the organ. The minimization of the network volume is subjected to several other requirements among which the following biophysical conditions are considered:
 (**I**). Total bulk volume of the organ, regardless of organ form.
 (**II**). Inlet and outlet features of the networks of the organ (ureter, main bronchus, etc).
(**III**). Sharing of space among the organs within the thorax and abdomen and special anatomic features needed to move materials through the organ, such as the diaphragm.
It will be shown that conditions (**I**) and (**II**), together with some elements of (**III**), are sufficient to derive an approximation to the kidney form and to the overall exterior form of both lungs above the diaphragm. The kidney form is simpler in the sense that it is not as time dependent as the lung is during the breathing cycle. Since the total volume of a kidney is small, the kidney is only partially conditioned by (**III**). Therefore the kidney would be the first testing ground for the concepts and techniques of the model. Naturally, a full accounting of condition (**III**) would require a very complex set of calculations, a goal that is outside the scope of this paper.
A common use of variational techniques is found in the following classic isoperimetric theorem: for a given volume value, the geometric form with the least area is the sphere, which is also the form containing the maximum volume for a given area. In this model it is similarly discovered that the least network volume for a given organ volume is found in the modified ellipsoid organ form, which also corresponds to the maximum organ volume for a given network volume.
The model is based on the analysis of distribution networks with fractal characteristics presented in [2], where complete details and additional references on related issues are given. The initial point of the model is a formula (Eq. (8) of [2]) for the calculation of the volume of a network of fluid distribution,

$$V_{net} = (C_{01}F_A A_0 / V) \int X(\mathbf{r}) dV \equiv (P/V) I_X \qquad (1)$$

where: $P = C_{01}F_A A_0$; $C_{01}$ is a constant; $F_A$ is a function of other parameters of the network; $A_0$ is the cross section area of the main branch of the network such as the renal artery; $V$ is the organ volume; and $I_X$ is the integral over the organ volume region of the circulation distance function $X(\mathbf{r})$. This is the distance inside the network between the position $\mathbf{r} = (r_x, r_y, r_z)$ of a site of the organ, such as a nephron, to the average inlet position of the network. In this model, it is assumed that the most relevant effect of the organ form on the network volume is contributed by $I_X$, which is the quantity to be minimized, while leaving $P$ and $V$ constant in this process.

To model circulatory organs such as the kidney and lung, we should consider the three main networks: the arterial, the venous, and the urine drainage in the kidney and the bronchial in the lung. Due to the similarities and geometric parallelism of the venous and arterial networks within these organs, these two networks are summarized into a single network. In this modeling approximation, the organ contains two major networks: $A$ and $B$. In the case of the kidney these are the urine drainage network ($A$) and the combined arterial and venous network ($B$). In the lung a similar association is to have the bronchial network ($A$), and the blood network ($B$).

Thus by extending Eq. (1) to the total volume of network $A$ and $B$,

$$V_{netAB} = (P_A I_{XA} + P_B I_{XB}) / V \qquad (2)$$

To describe further the various network geometric functions a conventional spherical coordinate system is introduced, $(r, \theta, \varphi) = (r, \Omega)$. The origin of the coordinate is placed at the origin of network $A$ and the origin of network $B$ is placed at a position $c_B$ on the $+r_z$ axis (Fig. 1). The inlet/outlet connections of the organ are in the region of the $-r_x$ axis. The circulation distance function $X_A(r, \theta, \varphi)$ is in general given by (Eq. (7) of [2]), $X_A = f_{A1}r^u + f_{A2}r^v \chi_A(\theta, \varphi)$, where, $f_{A1} > 1$, $f_{A1} >> f_{A2}$ and similarly for $X_B$. In [2] it is shown that the first term of $X_A$ and $X_B$, with $u = 1$ (the "Euclidean" distance) is the dominant contribution, therefore this is the only term under consideration in what follows. Then, $X_A = f_{A1}r$, and $X_B = f_{B1}r_B$, where $r_B$ is the distance between $(r_x, r_y, r_z)$ and $(0, 0, c_B)$.

The region of integration for $r_x > 0$ is exteriorly delimited by the organ form represented by the surface $R(\theta, \varphi)$, (Fig. 1) which will be determined as a solution of the equation resulting from the minimization of the network volume. Going from $r_x > 0$ to $r_x < 0$ the region of integration near the origin of the networks (Fig. 1) can include a small region delimited by the $R_{in}(\theta, \varphi)$ surface needed to provide space to connect the organ to the major (inlet or outlet) vessels of body circulation (such as renal artery, vein and ureter). The function $R_{in}$ is treated as a given surface that is a boundary condition for $R$. In this way condition (**II**) listed in the introduction is accounted for. The union of $R_{in}$ and $R$ surfaces represent the whole organ form, and using them in Eq. (2) the network volume is

$$V_{netAB} = V^{-1} \int_{\Omega'} \int_{R_{in}(\Omega)}^{R(\Omega)} (Q_A r + Q_B r_B) r^2 dr d\Omega \equiv V^{-1} I(R) \qquad (3)$$

where, $Q_A = P_A f_{A1}$, $Q_B = P_B f_{B1}$, and $d\Omega = \sin\varphi d\varphi d\theta$. A basic trigonometric application to the distance $r_B$ shows that, $r_B(r, \varphi) = (r^2 + c_B^2 - 2rc_B\cos\varphi)^{1/2}$. Also using the same variables and functions the volume of the organ is

$$V = V(R) = \int_{\Omega'} \int_{R_{in}(\Omega)}^{R(\Omega)} r^2 dr d\Omega \qquad (4)$$

To minimize the networks volume under the variations of the organ form $R(\Omega)$ the integral $I(R)$ (Eq. (3)) is regarded as a functional of $R(\Omega)$. To incorporate condition (**I**) of the introduction, the minimization is done for a given value of the organ volume, $V_o = V(R)$. A Lagrange multiplier $\lambda$ is introduced to combine (3) and (4), and determine the extrema in relation to $R$ of the auxiliary functional $V_{aux}(R)$

$$V_{aux}(R) = V_o^{-1} I(R) + \lambda(V_o - V(R)) \equiv \int_{\Omega'} F_{aux}(R) d\Omega + \lambda V_o \quad (5)$$

where $F_{aux}(R)$ is defined by (5). Since $F_{aux}(R)$ contains no derivatives of $R$, the problem is reduced to equating to zero the partial derivative of $F_{aux}(R)$ in relation to $R$, which determines the following equation for $R$,

$$\partial F_{aux} / \partial R = [(Q_A R + Q_B r_B(R, \varphi))/V_o - \lambda] R^2 = 0 \quad (6)$$

The analogy to the isoperimetric theorems is established as follows. If we seek the largest organ volume for a given network volume, $V_{ABo} = V_{netAB}$, then from Eq. (3), $V(R) = I(R)/V_{ABo}$. This last equation is combined with Eq. (4) with the help of a new Lagrange's multiplier $\lambda'$ in order to establish a new auxiliary functional, $V'_{aux}(R) = V(R) + \lambda'[(I(R)/V_{ABo}) - V(R)]$, which has an extrema in relation to $R$ determined by the same Eq. (6), with the identification, $V_o \lambda = V_{ABo}(\lambda' - 1)/\lambda'$.

Eq. (6) is a simple quadratic equation for R. The solution formula with the square root in the denominator is chosen. Rewriting Eq. (6) as, $aR^2 + 2bR - 2c = 0$, then, $R(\varphi) = 2c/\{b \pm [b^2 + 2ac]^{1/2}\}$, where the extra factors of 2 are introduced to avoid a later tedious redefinition of parameters. In this case, $a = (Q_B^2 - Q_A^2)/\lambda Q_A V_o = (e_c/c_B) - (1/c_L)$; $b = 1 - e_c \cos\varphi$, and $c = (c_L - e_c c_B)/2$, where the new parameters are, $c_L = \lambda V_o/Q_A$ and $e_c = c_B Q_B^2/\lambda Q_A V_o$. Notice that the four original positive parameters $Q_A$, $Q_B$, $V_o$, $c_B$, together with $\lambda$ are reduced to the three effective parameters for $R(\varphi)$ which are $c_B$, $c_L$, and $e_c$. In terms of dimensional analysis, it turns out that $c_L$ is a length and $e_c$ is dimensionless, because initially, $\lambda$ is dimensionless, $c_B$ is a length and $Q_A$, $Q_B$ have dimension of area. The range of variation of some of these parameters is restricted, as is discussed below. Writing the solution in a more explicit form

$$R(\varphi) = 2c/\{1 - e_c \cos\varphi + [(1 - e_c \cos\varphi)^2 + g_e]^{1/2}\} \quad (7)$$

where, $g_e \equiv 2ac = -[e_c - (c_L/c_B)][e_c - (c_B/c_L)]$. This solution has axial symmetry ($r_z$ axis), but the imposition of the boundary condition with the $R_{in}(\theta, \varphi)$ surface would in general break the axial symmetry of the final solution.

In Eq. (7) the + root has been chosen, to allow the important particular case, $g_e = 0$ to be well defined, and continuous with $g_e \neq 0$. When $g_e = 0$ the solution (7) is an ellipsoid with eccentricity $e_c$, and the main axis on the $r_z$ axis. This would require that $0 \leq e_c < 1$, and $c \geq 0$ (or $c_L/c_B \geq e_c$). The case $g_e = 0$, correspond to two networks with the same volume characteristic, though not identical networks.

When $g_e \neq 0$, the solution is a modified ellipsoid similar to an egg form (Fig. 2) which is named here as the g-ellipsoid (no previous identification of this surface was found). The g-ellipsoid is made by rotating the g-ellipse around the $r_z$ axis. The parameter, $g_e = 2ac \geq -(1 - e_c)^2$, makes the square root of the denominator of Eq. (7) always well defined.

The original meaning of $e_c$ as the eccentricity is formally lost when $g_e \neq 0$. In fact for $e_c > 1$, still there is a g-ellipse ($g_e > 0$) that becomes closer to a circle as $e_c$ becomes larger, as shown in Fig 2 most exterior curve. However, many g-ellipses are close to a single ellipse, especially when $g_e$ is small (Fig. 2). A much better approximation of the g-ellipse can be obtained with two different ellipses along the $r_z$ axis. In Fig. 2. the most interior ellipse of the illustration follows a g-ellipse ($g_e = -0.109$) very closely for $r_z < 1$, similarly another half ellipse of larger eccentricity closely follows the other side of this g-ellipse ($r_z > 1.2$). This double ellipse approximation simplifies a great deal the applications of the g-ellipse discussed below. In this approximation the g-ellipse becomes a "natural interpolation" of two half ellipses of similar minor diameter with different major diameter.

To have a model for $R(\theta, \varphi)$ even closer to real organ forms some elements from condition (**III**) must be included. The first basic element to consider is the asymmetric restrictions on the available space to the organ along the $x, y,$ and $z$ directions. We can incorporate this into the arguments leading to Eq. (7) by considering only an angular slice, between $\theta$ and $\theta + \Delta\theta$, of the total organ volume as the region of integration. Then the parameters $\lambda$ and $V_o$ are replaced by functions of $\theta$. The new solution for $R$ is still given by Eq. (7), with new parameters $c, e_c$ and $g_e$ which are functions of $\theta,$ which in turn depend on the way the additional boundary conditions are imposed. This last argument shows that a more complete solution for $R$ can be found within a larger family of surfaces generated by g-ellipses. Such solution is constructed by a more general g-ellipsoid that has, in the double ellipsoid approximation, different eccentricities along the $x, y, z$ directions. Such surface is here called the g3-ellipsoid.

The use of the g3-ellipsoid and ellipsoid in modeling the kidney and the overall exterior form of both lungs is shown in Fig. 3 and 4. The fitness of the g-ellipse can be seen by looking at the cross sections and the two-dimensional projections of the organ forms. Other exterior views of the kidney that are not shown, display a similar fitness to the g-ellipse, as shown in Fig. 3. These illustrations are not intended as rigorous evaluations, rather as examples of possible applications of the model developed so far. No statistical analysis of these results has been attempted, an issue that by itself would require some theoretical development. Choosing from organ images as close to reality as possible, a selection was made on the basis of the overall visual clarity of the final picture. There is plenty of medical imaging of the lung of normal human subjects, where the type of curve fitting shown in Fig. 4 is a very common result, but it is still a better illustration a traditional anatomical drawing, as is shown. Fig. 4 suggests that the exterior form of both lungs follow a single optimal organ design given by a section of an ellipsoid (or g3-ellipsoid). It would be of great interest to elaborate further this description of the lungs at the various stages of the breathing cycle.

An important and arduous task is the modeling of the surface $R_{in}(\theta, \varphi)$ and its detailed relation to the g3-ellipsoid. This is a first step towards a more complete inclusion of condition (**III**). These, and many other modeling issues about the organs are now available to quantitative analysis through the possible applications and extensions of this model.

**Acknowledgments.** Marie-Claude Vuille's assistance, discussion and encouragement have been invaluable in the preparation of this article and its expanded version.

FIG. 1. Schematics of coordinates and geometric variables of model.

FIG. 2. Ellipses ($g_e = 0$), and g-ellipses compared. From Eq. (7), the parameters of the five closed curves from inside to outside are:
$c$ = 0.96 ; 0.933; 1.0 ; 1.07 ; 1.833
$e_c$ = 0.68 ; 0.65 ; 0.75 ; 0.85 ; 2.0
$g_e$ = 0.0 ; –0.109; 0.0 ; 0.124; 2.674
The cut-off curve ($r_x > 0$) is the ellipse: $(r_x/1.355)^2 + ((r_z-1.3)/2.7)^2 = 1$; (ecc. = 0.865).

FIG. 3. Photographs of sheep's kidneys, approximated by an ellipse (a), and by a g-ellipse (b) using the method of the double ellipse. The dashed straight lines show the axis directions of the ellipses.

FIG. 4. Human lung forms approximated by an ellipsoid (dashed curves). Part (a), in situ illustration of thorax, from [5]. Part (b), is a transverse section of the thorax, from [6].

\MIOF\MIF3BR

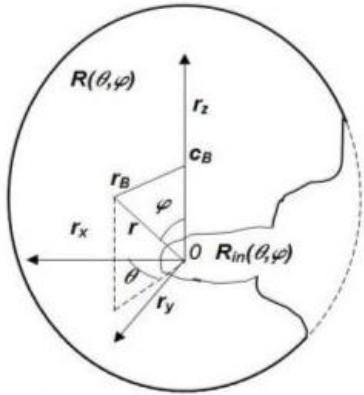

FIG. 1

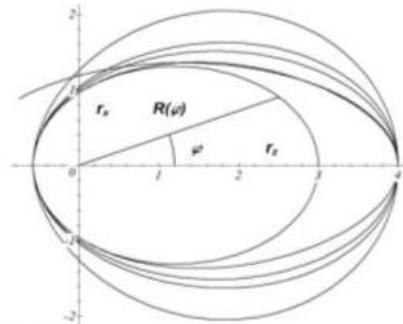

FIG. 2

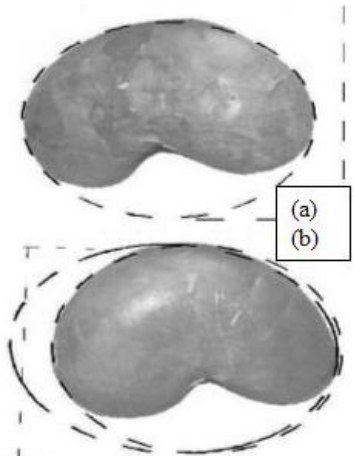

FIG. 3

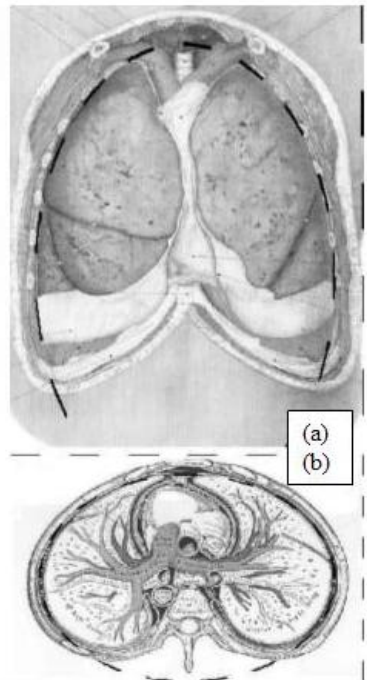

FIG. 4

Preprint May 2005
\wagu\MIOF\MIF3BP